# Time Dependent Clustering Analysis of the
# Second BATSE Gamma-Ray Burst Catalog

J. J. Brainerd,[1]*

C. A. Meegan,[2] Michael S. Briggs,[1] G. N. Pendleton,[1] M. N. Brock[2]

[1]University of Alabama in Huntsville

[2]NASA/Marshall Space Flight Center

December 6, 1994

## ABSTRACT

A time dependent two-point correlation-function analysis of the BATSE 2B catalog finds no evidence of burst repetition. As part of this analysis, we discuss the effects of sky exposure on the observability of burst repetition and present the equation describing the signature of burst repetition in the data. For a model of all burst repetition from a source occurring in less than five days we derive upper limits on the number of bursts in the catalog from repeaters and model-dependent upper limits on the fraction of burst sources that produce multiple outbursts.

*Subject Headings*: Gamma-Rays: Bursts

---

* Mailing address: Space Sciences Lab, ES-84, NASA/Marshall Space Flight Center, Huntsville, AL 35812; E-Mail: brainerd@ssl.msfc.nasa.gov

## 1. INTRODUCTION

Gamma-ray burst observations by the Burst and Transient Source Experiment (BATSE) on the Compton Gamma-Ray Observatory (CGRO) have tightened the geometric constraints on galactic gamma-ray burst models to the point that an extragalactic origin now provides the most natural explanation for the gamma-ray burst spatial distribution (Briggs *et al.* 1994; Hakkila *et al.* 1994). The question that now must be answered is what useful constraints can be placed on models of extragalactic gamma-ray bursts by the observations. One of the more important constraints is on the rate at which sources produce gamma-ray bursts. If burst repetition were proven, then destructive burst models such as the merging neutron star model would be disproven and other models would have limits placed on their energy production mechanisms.

If gamma-ray burst sources repeat, one expects them to produce small scale clustering, with the clustering scale set by the burst location error. Several authors have analyzed with mixed results the 1B Gamma-Ray Burst Catalog for small scale clustering (Blumenthal *et al.* 1994; Quashnock & Lamb 1993). A subsequent analysis of the 2B catalog finds no evidence of burst clustering or repetition with either a two-point correlation-function analysis or a nearest neighbor analysis and places an upper limit of approximately 20% on the fraction of bursts observed as repeater bursts (Meegan *et al.* 1994).

Wang and Lingenfelter have conducted a time-dependent analysis of the 1B catalog (Wang & Lingenfelter 1993, 1994). The first article argues that five bursts with overlapping location errors are from a single source. The second article presents the results of a time dependent two-point correlation-function analysis of the 1B catalog. They find an excess of burst pairs with separations of $t < 5$ days and $\theta < 4°$ which they assert is statistically significant and evidence of burst repetition.

In this paper we perform a time dependent two-point correlation-function analysis of the 2B catalog (Meegan *et al.* 1994) on both the full catalog of 585 bursts and the subset of 485 bursts with non-MAXBC locations (defined in §2). In §2 we discuss the effect of earth blockage, trigger off-time, and data loss on the ability of BATSE to observe burst repetition. The polynomial time variable $\tau$ defined by Wang and Lingenfelter and the effects of a variable burst detection rate are discussed in §3. We discuss the expected behavior of the two-point correlation-function in §4, and we give our results from such an analysis of the 2B catalog in §5.

## 2. DETECTING REPEATING BURSTS

The number of bursts that BATSE can detect from a repeating source is dependent on the fraction of the sky occulted by the Earth and the fraction of time the burst trigger is enabled. These effects are easily modeled if one assumes that the interval of time between outbursts is long compared to the orbital period of the spacecraft, so that the probability of observing each outburst from a source is uncorrelated with the observation of previous outbursts. Generally a repeater must produce many bursts above threshold for BATSE to observe two or more bursts from the same source. A consequence of this is that the observations set stronger constraints on the repetition models with high intrinsic burst-rates than those with low intrinsic burst-rates.

Because the CGRO tape recorders failed in the second year of operation, a significant fraction of bursts in the 2B catalog only have locations determined from the MAXBC data type—the maximum count rate over 1 second in each detector module in the 50 keV to 300 keV energy band. In contrast, all bursts in the 1B catalog have non-MAXBC locations. The MAXBC data type is the most problematic BATSE data type for locations, producing locations with larger errors than the other BATSE data types. Exclusion of MAXBC positions from a sample improves the overall location error at the expense of a smaller trigger enable time. Meegan *et al.* (1994) find that the signal of repetition in the two-point correlation-function is proportional to the sky exposure regardless of the underlying repetition model. Therefore, if MAXBC locations are excluded, a signal for burst repetition would be $\approx 25\%$ smaller in the 2B catalog than in the 1B catalog. This loss is almost precisely compensated for by the better statistics from the larger sample size.

## 3. DISTRIBUTION OF BURST TIME SEPARATIONS

If bursts were detected at a uniform rate over a time interval $T$, the time separation $\Delta t$ of pairs of bursts would be uniformly distributed in the variable

$$\tau = 1 - \left(1 - \frac{\Delta t}{T}\right)^2 , \qquad (1)$$

(Wang and Lingenfelter 1994). For $N$ bursts, the average number of burst pairs per unit $\tau$ is $N(N-1)/2$ if the instrument is always on and if the burst trigger threshold is constant; as discussed below, both of these assumptions are false, but for the present discussion we accept them as true.

A uniform distribution of burst pairs with $\tau$ does not imply a normal distribution in $\tau$. In fact, the distribution of burst pairs in $\tau$ is highly correlated, and this correlation is different for different intervals in $\tau$. For instance, if the interval $0 < \tau < 1$ is divided into $n$ subintervals of equal size, then the number of burst pairs in the subinterval $0 < \tau < 1/n$ for a particular burst decreases by one and increases by an arbitrary amount over the number of pairs in this interval for the preceding burst. On the other hand, in the interval $(n-1)/n < \tau < 1$ the number of burst pairs for a particular burst never increases but can decrease by an arbitrary amount from the number of pairs for the preceding burst. The statistics of the two bins are therefore different.

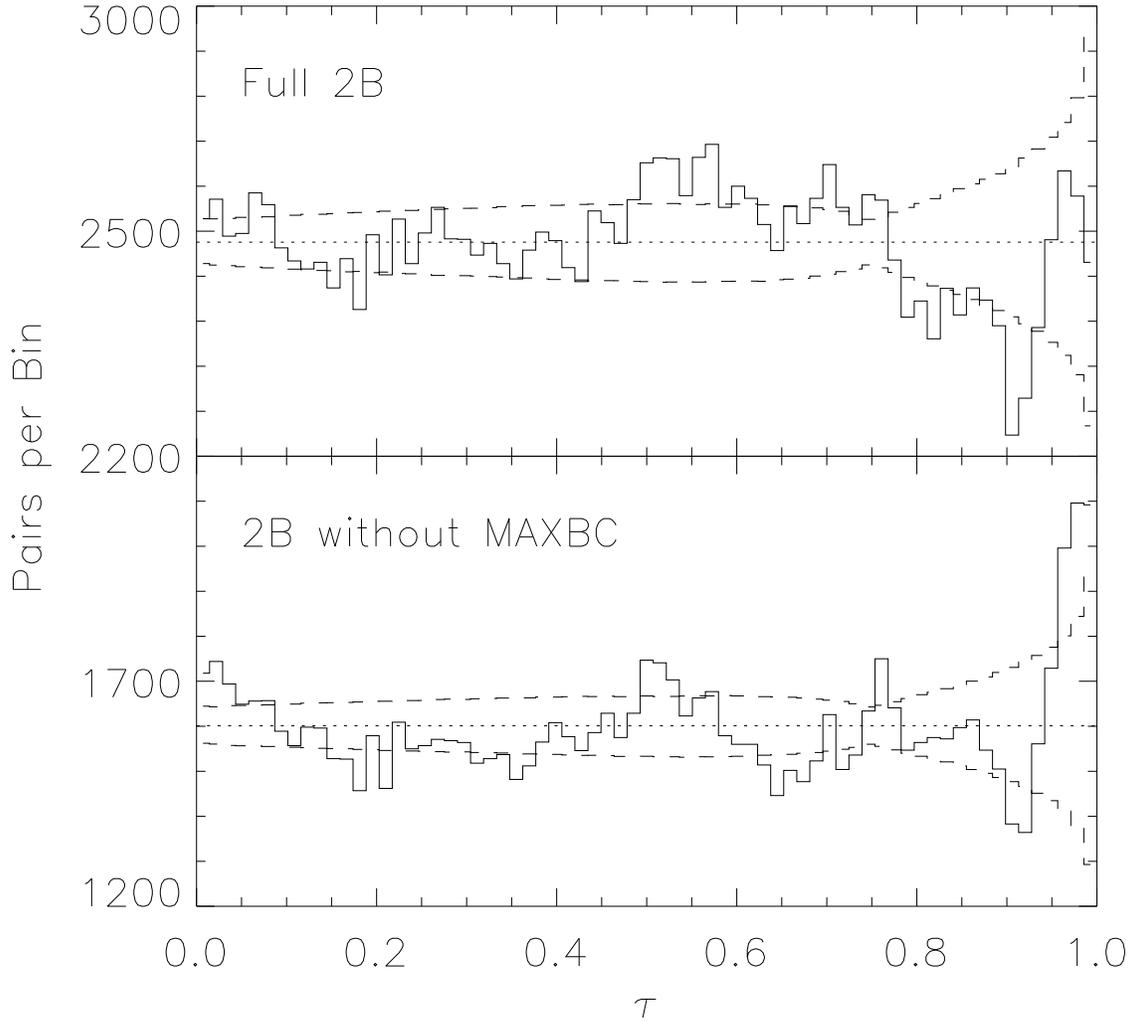

Figure 1. Number of bursts in each $\tau$ bin. The data are given by the solid line histograms. The data for the full 2B catalog are plotted in the upper diagram and the data for the subset of bursts with non-MAXBC locations are plotted in the lower diagram. The expected averages are given by the dotted curves and the $1\sigma$ statistical deviations are given by the dashed curves.

Distributions in $\tau$ for burst pairs in the 2B catalog are given in Figure 1. The upper figure gives the distribution of burst pairs for the full 2B catalog; the lower figure gives the distribution for the subset of the 2B catalog with non-MAXBC locations. These histograms have 69 bins covering 691 days, and so the first bin covers 5.00 days. This division of the 2B catalog in $\tau$ is chosen because the earlier analysis of Wang and Lingenfelter found a deviation from the average expected value for an interval of 0 to 5 days in the 1B catalog. The data is plotted as the solid line histogram while the $1\sigma$ statistical deviations above and below the average as derived from Monte Carlo simulation are plotted as dotted histograms. A normal distribution gives $\sigma = 50$ in the upper plot and $\sigma = 41$ in the lower plot. The deviations from a normal distribution are substantial, particularly for large $\tau$.

The rate at which BATSE triggers on bursts is highly variable in time due to several factors. First, the trigger is disabled during intervals of the orbit where particle precipitation is a problem. Second, the burst trigger threshold is raised substantially for 90 minutes after a trigger, which lowers the detection rate of gamma-ray bursts. Third, detector thresholds were changed in response to the transient source GRO J0422+32. Fourth, the fall in the solar flare rate with the approach of the solar minimum lowered the fraction of time the trigger threshold was raised. Finally, excluding bursts with MAXBC locations from the analysis lowers the the burst detection rate for the second year of the 2B catalog. The variable burst detection rate makes the average number of bursts expected per unit $\tau$ variable. For example, if the detection rate increases linearly so that it is 20% larger at the end of the catalog than at the beginning of the catalog, the distribution in $\tau$ will be above average by 0.9% for $\tau = 0$, and it will be below average by 2.2% for $\tau = 1$. Both of these deviations are smaller than the standard deviations of 2% at $\tau = 0$ and 12% at $\tau = 1$. Because the actual change in the detection rate is of this order, rising from an average of 0.81 to 0.88 bursts per day for the full 2B catalog, and falling from 0.81 to 0.60 bursts per day for the non-MAXBC located subset, one sees that the effects of a variable burst detection rate are smaller than the $\tau$-dependent statistical fluctuations.

## 4. THEORETICAL EXPECTATIONS

The isotropic repeater model has $m$ repeater sources with each source $i$ producing $\nu_i$ observed outbursts. The distribution of $\nu_i$ outbursts on the sky around the source position as a consequence of burst location errors is given by

$$\frac{dP_i}{d\Omega} = \frac{\nu_i}{2\pi\mu\left(1 - e^{-2/\mu}\right)}\, e^{-\frac{1-\cos\theta}{\mu}}\,, \tag{2}$$

where $\mu$ is the standard deviation in $1 - \cos\theta$, and is assumed to be the same for all bursts. For $\mu \ll 1$, the position angle error is $\sigma = \sqrt{2\mu}$. The two-point correlation-function $w(\theta)$ can be solved analytically for this distribution function. Averaged over the interval

$x_0 < \cos\theta < x_1$, one has

$$\langle w(\theta) \rangle_{x_0, x_1} = S_{\tau_0, \tau_1} \left\{ \frac{4e^{-\frac{2}{\mu}} \left[ \cosh\left( \frac{\sqrt{2+2x_1}}{\mu} \right) - \cosh\left( \frac{\sqrt{2+2x_0}}{\mu} \right) \right]}{(x_1 - x_0)\left(1 - e^{-2/\mu}\right)^2} - 1 \right\}, \qquad (3)$$

where $S_{\tau_0, \tau_1} = \sum_{i=1}^{m} \nu_i (\nu_i - 1) / N(N-1)$, with $N$ the total number of bursts in the sample.

To apply equation (3) to clustering over the time delay interval $\tau_0$ to $\tau_1$, we must change $S_{\tau_0, \tau_1}$ to

$$S_{\tau_0, \tau_1} = \frac{\sum_{i=1}^{m} \sum_{j=1}^{\nu_i} n_{i,j}(\tau_0, \tau_1)}{\sum_{i=1}^{N} N_i(\tau_0, \tau_1)}, \qquad (4)$$

where $\nu_{i,j}(\tau_0, \tau_1)$ is the average number of bursts from source $i$ that fall between $\tau_0$ and $\tau_1$ after burst $j$ and $N_i(\tau_0, \tau_1)$ is the average number of bursts that follow gamma-ray burst $i$ by $\tau_0 < \tau < \tau_1$. Both $\nu_{i,j}(\tau_0, \tau_1)$ and $N_i(\tau_0, \tau_1)$ are highly model dependent.

If the time in which a repeater produces all of its bursts is shorter than the time covered by the first bin in $\tau$, then $S_{\tau_0, \tau_1} = 0$ for all bins but the first. In the first bin, the numerator will be $\sum_{i=1}^{m} \nu_i (\nu_i - 1) / 2$. If for simplicity we assume that the contribution of the repeaters is negligible to the term in the denominator, then the denominator is $\sum_{i=1}^{N} N_i(\tau_0, \tau_1) \approx \Delta\tau N(N-1) / 2$. Therefore in the first bin $S_{\tau_0, \tau_1}$ is larger than the value found for the full catalog by a factor of $1/\Delta\tau$. Because the statistical fluctuations in the first bin vary approximately as the square root of the number of burst pairs, the signal-to-noise ratio of this signature will rise.

On the other hand, if the bursts from a repeater are randomly distributed, then both the denominator and numerator of $S_{\tau_0, \tau_1}$ are equal to the values for the full catalog value times $\Delta\tau_i$, and so the equation for $w(\theta)$ is unchanged. But the statistical noise is worse than for the full catalog, so subdividing over $\tau$ decreases the odds of finding clustering.

These two models are the extremes of behavior. A third type of behavior is for all repeaters to have the same period so that periodically spaced bins contain unusually large numbers of pairs of repeater bursts. If $m$ repeaters each produce $\nu$ bursts that are periodically spaced over a time scale long compared to the width of the time bins, then the numerator of equation (6) is $m(\nu - 1)$ for the bins spanning repetition times, and, as with the first example, the signal of repetition increases as the number of time bins increases. But the size of the signal is smaller by a factor of $\nu$ from the first example.

The best hope of seeing a clustering in a time dependent analysis is for it to be in the first time interval.

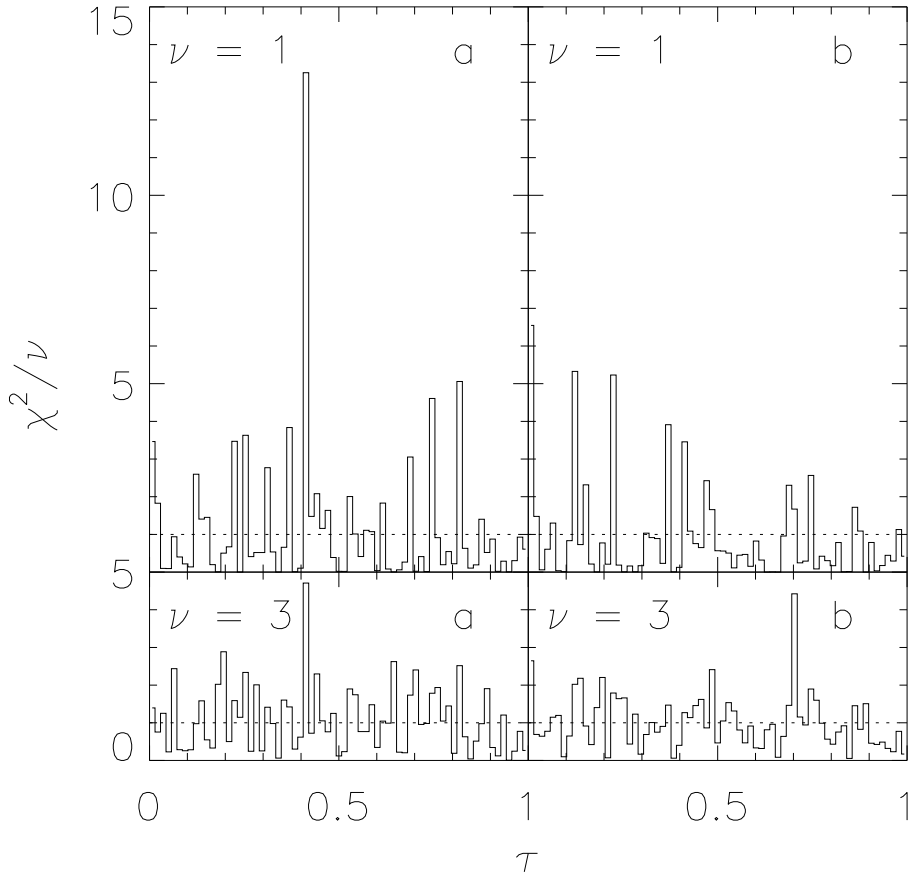

Figure 2. Histogram of reduced $\chi^2$ for (a) the full 2B catalog of 585 bursts and (b) the non-MAXBC subset of 485 bursts from the 2B catalog. The upper diagram gives the values of $\chi^2$ for the first bin in $\cos\theta$ ($n = 1$). The lower diagram gives $\chi^2/3$ for the first three bins in $\cos\theta$ ($n = 3$).

## 5.   ANALYSIS OF THE 2B CATALOG

Because the statistical fluctuations in the number of burst pairs in an interval $\Delta\tau$ is dependent on $\tau$, the distribution of burst pairs in two dimensional intervals of $\Delta(\cos\theta)$ and $\Delta\tau$ is highly correlated and not described by a Poisson distribution. However, the distribution over $\cos\theta$ for a particular value of $\tau$ is Poisson distributed. As a consequence, the significance of a particular number of bursts in an interval of $\Delta(\cos\theta)$ and $\Delta\tau$ should be determined from the total number of burst pairs in this interval of $\Delta\tau$.

The choice of bin size in $\cos\theta$ is determined by the value of $\mu$ in equation (4) set by the BATSE location errors. These errors have a 4° systematic error and a statistical error dependent on the fluence of the burst. Setting $\sigma = 6°$, which describes the error distribution for 85% of bursts, gives $\mu = 0.0055$. Then a choice of 400 bins over the interval $[-1, 1]$ for $\cos\theta$ gives a value of $\langle w(\theta) \rangle_{0.990,0.995} / \langle w(\theta) \rangle_{0.995,1} \approx 0.63$. Successive bins fall in value by this factor, so the first three bins contain 83% of the signal. The fourth bin only contains 7% of the signal.

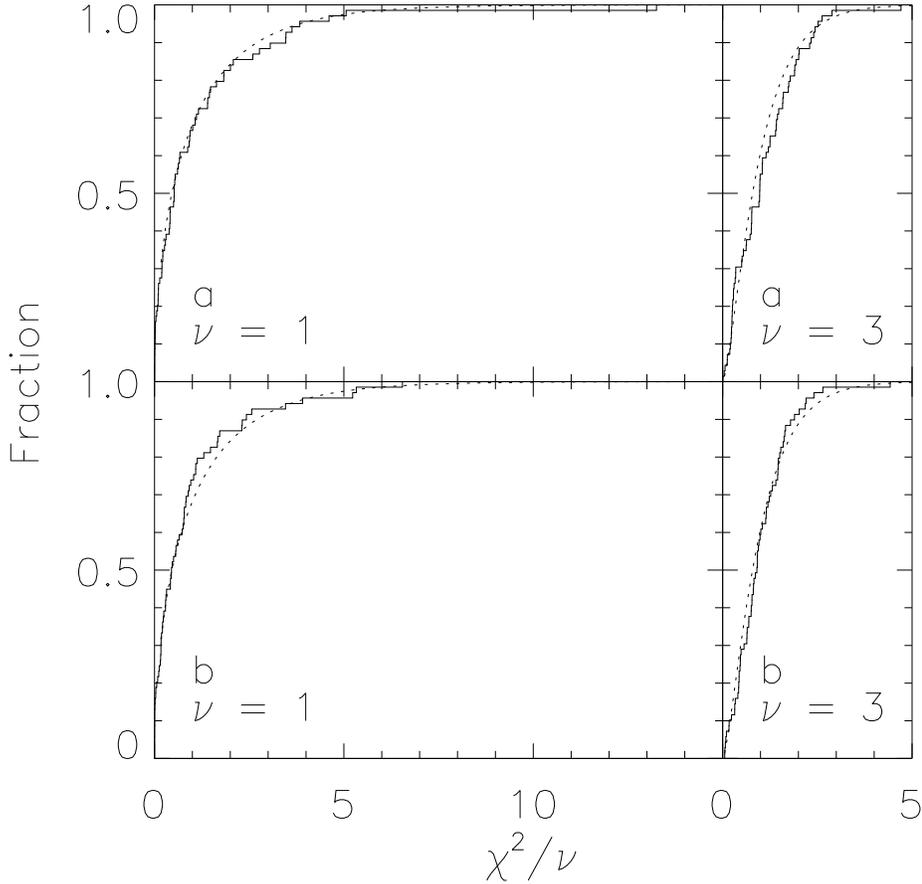

Figure 3. Cumulative distribution of reduced $\chi^2$ for the 2B catalog. Labels (a) and (b) are for the data sets described in Fig. 2. The dotted lines give the reduced $\chi^2$ distribution functions.

Figure 2a shows for the full 2B catalog the reduced $\chi^2$ of the first ($n = 1$) and the first three ($n = 3$) bins in $\cos\theta$ for each interval in $\tau$. The same is shown for the subset of 2B bursts with non-MAXBC locations in Figure 2b. The cumulative distribution of the reduced $\chi^2$ of the data is plotted in Figure 3 along with the reduced $\chi^2$ cumulative distribution function. These plots show that only one bin is unusually high, and that is the first bin in $\cos\theta$ covering the interval $0.406 < \tau < 0.420$ ($158.3$ days $< \Delta t < 164.9$ days) for the full catalog. For 69 bins one expects 0.02 bins to have a fluctuation of this magnitude. But the first three bins in $\cos\theta$ for this time has a reduced $\chi^2$ of 4.7, which is expected for 0.2 bins in a sample of 69 bins. Additional evidence of the statistical origin of the high bin in the upper plot of Figure 1a is that adjacent bins, the bin at a multiple of 2 from the high bin ($0.710 < \tau < 0.725$), and the bins at $0.406 < \tau < 0.420$ in Figure 2b have values expected for an isotropic distribution. The bin covering $0 < \tau < 0.014$ ($0 < \Delta t < 5.0$ days) does not have an exceptionally large deviation in any of the figures. Its largest deviation is for the first bin in $\cos\theta$ in Figure 2b, for which this bin has the largest fluctuation above average of all the bins. But one expects to have 0.7 bins with this large of a fluctuation in a sample of 69 bins.

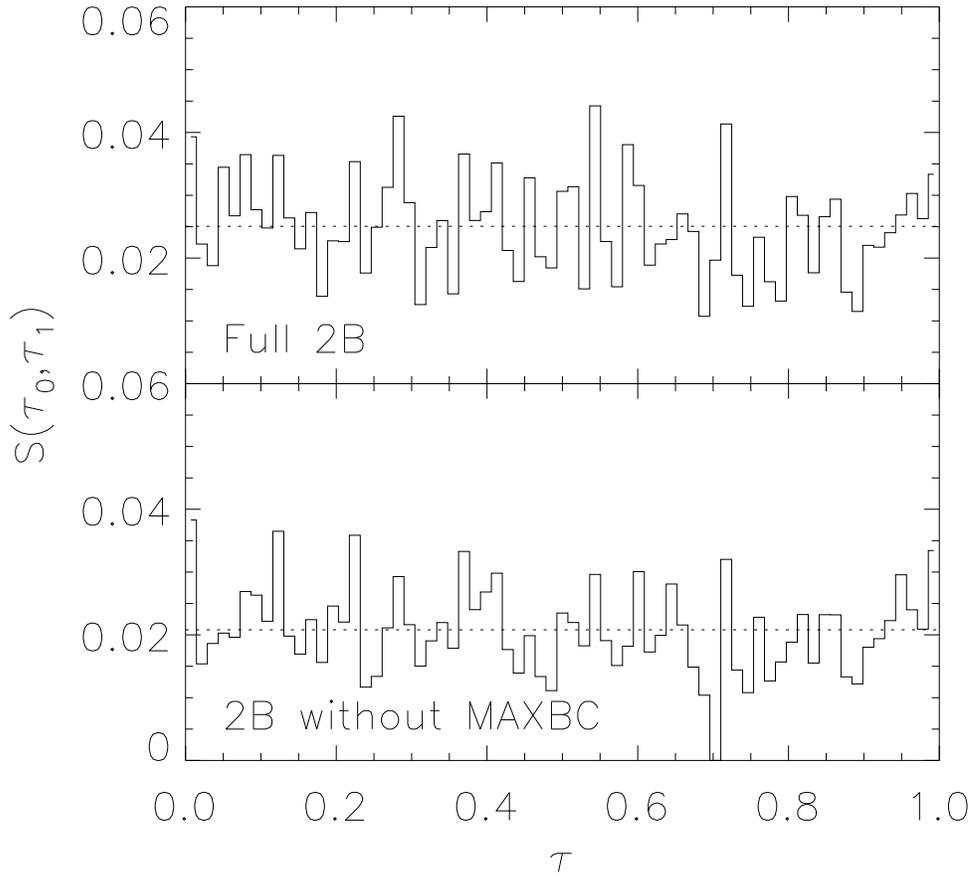

Figure 4. Limits on $S(\tau_0, \tau_1)$ of equation (4). At these limits the probability of a larger reduced $\chi^2$ for the first three bins in $\cos\theta$ is 1%. The bin in the lower diagram that goes to zero has a reduced $\chi^2$ that is below this probability at the $\chi^2$ minimum.

Limits on the value of $S_{\tau_0,\tau_1}$ are plotted in Figure 4 for each $\tau$ bin. These limits are found by applying a modified version of equation (3) to the first three bins in $\cos\theta$ and finding the value of $S_{\tau_0,\tau_1}$ that produces a $\chi^2$ of 1% probability. The error model we use is the sum of two Fisher distributions, one with a $6°$ error and normalized to account for 85% of the bursts, the other with an error of $12°$ and accounting for the remaining 15% of the bursts. This model is derived from the error distribution of the 2B catalog. The bin at $\tau = 0.7$, which is set to zero, has a $\chi^2$ probability of less than 1% at the minimum value of $\chi^2$. The average values of $S_{\tau_0,\tau_1}$ are 0.025 for the full catalog and 0.021 for the subset without MAXBC locations. The values of $S_{\tau_0,\tau_1}$ in the first bin are 0.039 for the full catalog and 0.038 for the burst subset.

The full catalog produces a total of 2528 burst pairs over the first time bin, so for a model in which all observed repetitions from a source occurs within 5 days of the first burst, the upper limit on the number of sources appearing in the full catalog as repeaters is $m < 197/\nu(\nu-1)$ at the 99% confidence level, where we have assumed that each identifiable repeating source produced $\nu$ observed bursts. For $\nu = 3$, this implies that the fraction of observed bursts that are from identifiable repeaters is $f = m\nu/N < 0.17$. For $\nu = 2$ this limit rises to $f < 0.34$. The non-MAXBC located bursts produce 1818 burst pairs in the first time bin, which gives a limit on identifiable repeater sources of $m < 138/\nu(\nu-1)$ at the 99% confidence level. For $\nu = 3$ one has $f < 0.14$, and for $\nu = 2$, $f < 0.28$. From these limits on the data one can derive model dependent limits on the fraction of burst sources producing multiple observed and unobserved bursts. For repeaters producing 2, 3, 4, 6, 8, and 10 outbursts the upper limits on the fraction of repeating sources in the full catalog are 85%, 22%, 9.8%, 3.6%, 1.8%, and 1.1%, with the average number of observed bursts from identifiable repeaters ranging from 2 to 3.9. For the non-MAXBC located bursts, the 2 outburst repeater model has a formal upper limit on the repeater source fraction of 103% at the 99% confidence level, and therefore no physical upper limit, and the models with 3, 4, 6, 8, and 10 outbursts have upper limits on the repeater source fraction of 25%, 11%, 4.0%, 2.1%, and 1.3%, with the average number of bursts observed from identifiable repeaters ranging from 2.1 to 3.3.

We find no evidence of burst repetition in either the full 2B catalog or the subset of this catalog with non-MAXBC locations.

*Acknowledgements:* This work is supported by NASA contract NAG5-2056.